\documentstyle[12pt,epsf]{article}
\newlength{\abstractwidth}
\renewcommand{\thefootnote}{\fnsymbol{footnote}}
\renewcommand{\thanks}[1]{\footnote{#1}} 
\newcommand{\starttext}{
\setcounter{footnote}{0}
\renewcommand{\thefootnote}{\arabic{footnote}}}
\renewcommand{\theequation}{\thesection.\arabic{equation}}
\newcommand{\be}{\begin{equation}}
\newcommand{\bea}{\begin{eqnarray}}
\newcommand{\eea}{\end{eqnarray}}
\newcommand{\beq}{\begin{equation}}
\newcommand{\ee}{\end{equation}}
\newcommand{\eeq}{\end{equation}}

\newcommand{\<}{\langle}

\renewcommand{\>}{\rangle}
\def\ba{\begin{eqnarray}}
\def\ea{\end{eqnarray}}

\def\14{{1\over4}}
\def\12{{1 \over 2}}
\def\eq{&=&}

\def\h3{h^{3\over 2}}

\def\>{\rangle}
\def\<{\langle}

\def\des{de Sitter Space}

\def\cc{cosmological constant}

\def\sms{supermoduli--space}

\begin{document}
\renewcommand{\theequation}{\thesection.\arabic{equation}}
\begin{titlepage}
\bigskip
\rightline{SU-ITP 02-11} \rightline{hep-th/0204027}

\bigskip\bigskip\bigskip\bigskip

\centerline{\Large \bf {The Anthropic Landscape   }}
\centerline{\Large \bf {of String Theory }}

\bigskip\bigskip
\bigskip\bigskip

\centerline{\it L. Susskind  }
\medskip
\centerline{Department of Physics} \centerline{Stanford
University} \centerline{Stanford, CA 94305-4060}
\medskip
\medskip

\bigskip\bigskip
\begin{abstract}
In this lecture I make some educated guesses,  about the landscape
of  string theory vacua. Based on the recent work of a number of
authors, it seems plausible that the lanscape is unimaginably
large and diverse. Whether we like it or not, this is the kind of
behavior that gives credence to  the Anthropic Principle. I
discuss  the theoretical and conceptual  issues that arise in
developing a cosmology based on the diversity of environments
implicit in string theory.
\medskip
\noindent
\end{abstract}

\end{titlepage}
\starttext \baselineskip=18pt \setcounter{footnote}{0}

\setcounter{equation}{0}
\section{The Landscape }
The world--view shared by most physicists is that the laws of
nature are uniquely   described by some special action principle
that completely determines the vacuum, the spectrum of elementary
particles, the forces and the symmetries.  Experience with quantum
electrodynamics and quantum chromodynamics   suggests a world with a
small number of parameters and a unique ground state. For the most
part, string theorists bought into this paradigm. At first it was
hoped that string theory would be unique and explain the various
parameters that quantum field theory left unexplained. When this
turned out to be false, the belief developed that there were
exactly five string theories with names like type--2a and
Heterotic. This also turned out to  be wrong. Instead, a continuum
of theories were discovered that smoothly interpolated between the
five and also included a theory called M--Theory. The language
changed a little. One no longer spoke of different theories, but rather
different solutions of some master theory. The space of these
solutions is called {\it{The Moduli Space of Supersymmetric
Vacua}}. I will call it the \sms . Moving around on this \sms \ is
accomplished by varying certain dynamical $moduli$.  Examples of
moduli are the size and shape parameters of the compact internal
space that 4--dimensional  string theory always needs. These moduli are not
parameters in the theory but are more like fields. As you move
around in ordinary space, the moduli can vary and have their own
equations of motion. In a low energy approximation the moduli
appear as massless scalar fields. The beauty of the \sms \ point
of view is that there is only one theory but many solutions which
are
characterized by the values of the scalar field moduli. The
mathematics of the string theory is so precise that it is hard to
believe that there isn't a consistent mathematical framework
underlying the \sms \ vacua.

However the continuum of solutions in the \sms \ are all
supersymmetric with exact super--particle degeneracy  and
vanishing \cc . Furthermore they all have massless scalar
particles, the moduli themselves.  Obviously none of these vacua
can possibly be our world. Therefore the string theorist must
believe that there are other discrete islands lying off the coast
of the
 \sms . The hope now is that a single
non--supersymmetric island or at most a small number of islands
exist and that   non--supersymmetric physics will prove to be
approximately unique. This view is not inconsistent with present
knowledge (indeed it is possible that there are no such islands)
but I find it completely implausible. It is much more likely that
the number of discrete vacua is astronomical, measured not in the
millions or billions but in  googles or googleplexes \footnote{A
google is defined to be ten to the power one hundred. That is $
{\cal{G}}=10^{100}$. A googleplex is $10^{\cal{G}}$}.

This change in viewpoint is demanded by two facts, one
observational and one theoretical. The first is that
the expansion of the universe is accelerating. The simplest
explanation is a small but
non--zero cosmological constant. Evidently we have to expand our
thinking about  vacua  to include states
with non--zero vacuum energy. The incredible smallness and
apparent fine tuning of the \cc \ makes it absurdly improbable to
find a vacuum in the observed range
unless there are an enormous number of solutions with almost every
possible value of $\lambda$. It seems to me inevitable that if we
find one such vacuum we will find a huge number of them. I will
from now on call the space of all such
 string theory vacua the $landscape$.

The second fact is that some recent progress has been made in
exploring the landscape \cite{joe,andrei}. Before explaining the
new ideas I need to define more completely what I mean by the
landscape. The \sms \ is parameterized by the moduli which we can
think of as a collection of scalar fields $\Phi_n$. Unlike the
case of Goldstone bosons, points in the moduli space are not
related by a symmetry of the theory. Generically, in a quantum
field theory, changing the value of a non--Goldstone scalar involves
a change of potential energy. In other words   there is a non--zero field
potential $V(\Phi)$. Local minima of $V$ are what we call vacua.
If the local minimum is an absolute minimum the vacuum is stable.
Otherwise it is only metastable. The value of the potential
energy at the minimum is the \cc \ for that vacuum.

To the extent that the low energy properties of string theory can
be approximated by field theory, similar ideas apply. Bearing in
mind that the low energy approximation may break down in some
regions of the landscape, I will assume the existence of a set of
fields and a potential. The space of these fields is  the landscape.

The \sms \ is a special part of the landscape where the vacua are
supersymmetric and the potential $V(\Phi)$ is exactly zero. These
vacua are marginally stable and can be excited by giving the
moduli arbitrarily small time derivatives. On the \sms \ the \cc \
is also exactly zero. Roughly speaking, the \sms \ is a perfectly
flat plain at exactly zero altitude \footnote{By altitude I am of
course referring to the value of $V$.}. Once we move off the
plain, supersymmetry is broken and a non--zero potential
developes, usually through some non--perturbative mechanism. Thus
beyond the flat plain we encounter hills and valleys. We are
particularly interested in the valleys where we find local minima
of $V$. Each such minimum has its own vacuum energy.  The typical
value of the potential difference between neighboring valleys will
be some fraction of $M_p^4$ where $M_p$ is the Planck mass. The
potential barriers between minima will also be of similar height.
Thus if a vacuum is found with \cc \ of order $10^{-120}M_p^4$, it
will be surrounded by much higher hills and other valleys.

Next consider two large regions of space, each of which has the
scalars in some local minimum, the two minima being different.
 If the local minima are landscape--neighbors
then the two regions of space will be separated by a domain wall.
Inside the domain wall the scalars go over a ``mountain pass". The
interior of the regions are vacuum like with cosmological
constants. The domain wall which can also be called a membrane
has additional energy in the form of a membrane tension. Thus
there will be configurations of string theory which are not
globally described by a single vacuum but instead consists of many
domains separated by domain walls. Accordingly, the landscape in field space
is reflected in a complicated terrain in real space.

There are  scalar fields  that are not usually thought of as
moduli but once we leave the flat plain I don't think there is any
fundamental difference. These are the four--form field strengths
 first introduced in the context of the \cc \ by Brown
and Teitelboim \cite{claudio}. A simple analogy exists to help
visualize these fields and their potential. Think of 1+1
dimensional electrodynamics with electric fields\footnote{In
1+1 dimensions there is no magnetic field and the electric field
is a two form,
aka, a scalar density.}  $E$ and massive electrons. The electric field is
 constant in any region of space where there are no charges. The
 field energy is proportional to the square of the field
 strength. The electric field jumps by a quantized unit whenever
 an electron is passed. Going in one direction, say along the
 positive $x$ axis, the field makes a positive unit jump when an
 electron is passed and a negative jump when  a positron is passed. In
 this model different vacua are represented by different quantized
 values of the electric field while the electrons/positrons are the
 domain walls. The energy of a vacuum is proportional to $E^2.$
This model is not fundamentally different than the case with
scalar fields and a potential. In fact by bosonizing the theory it
can be expressed as a scalar field theory with
 a  potential
\be
V(\phi)= c\phi^2 +\mu cos\phi.
\label{boso}
\ee
If $\mu$ is not too small there are many minima representing the
different possible 2-form field strengths, each with a different
energy.

In 3+1 dimensions the corresponding construction requires a 4-form
field strength $F$ whose energy is also proportional to $F^2$.
This energy appears in the gravitational field equations as a
positive contribution to the \cc .
 The
analogue of the charged electrons are membranes which appear in
string theory and  function as domain walls to separate vacua
with different $F$. This theory can also be written in terms of a
scalar field with a  potential similar to  \ref{boso}. In the future I will
include such fields along with moduli  as coordinates of the
landscape.

Let's now consider a typical compactification of M--theory from
eleven
to 4 dimensions. The simplest example is gotten by choosing for the compact
directions a
7-torus. The torus has a number of moduli
representing the sizes and angles between the seven 1--cycles.
The 4--form fields have as their origin a 7--form field strength
\footnote{The fact that we have the number 7
appearing in two ways, as the number of compact dimensions and
as the number of indecies of the field strength is accidental.  }
which is one of the fundamental fields of M--theory. The 7--form
fields have
7 anti--symmetrized indecies. These non--vanishing 7--form can
be configured so that three of the indecies are identified with
compact dimensions and the remainder with uncompactified
spacetime. This can be done in  thirty five $=(7\times6\times5)/(1\times2\times3)$
 ways  which means that there are that many distinct
4--form fields in the uncompactified non--compact space.
More generally, in the kinds of compact manifolds
used in string theory to try to reproduce standard model physics
there can be hundreds of independent ways of ``wrapping" three compact
directions with flux, thus producing hundreds of 3+1 dimensional
 4--form fields. As in the case of 1+1 dimensional quantum
 electrodynamics, the field strengths are quantized, each in integer
 multiples of a basic field unit. A vacuum is specified by a set
 of integers $n_1,n_2,....,n_N$ where $N$ can be as big as several
 hundred or more. The  energy density of the energy of the 4--form fields
 has the form
 \be
\epsilon =\sum_{i=1}^N c_iN_i^2
 \ee
where the constants $c_i$ depend on the details of the compact
space.

The analogue of the electrons and positrons of the 1+1 dimensional
example are branes. The 11 dimensional M--theory has 5--branes
which fill 5 spatial directions and time. By wrapping
5--branes the same way the fluxes of the 4--forms   are wrapped on
 internal 3--cycles leaves 2--dimensional membranes in  3+1
 dimensions. These are the domain walls which separate different
 values of field strength. There are $N$ types of domain wall,
 each allowing a unit jump of one of the 4-forms.

 Bousso and Polchinski \cite{joe} begin by assuming they have located some
 deep minimum of the field potential at some point $\Phi_0$. The
 value of the potential is supposed to be very negative at this
 point, corresponding to a negative \cc , $\lambda_0$ of order the Planck
 scale. Also the 4-forms are assumed to vanish at this point.
  They then ask what kind of vacua can they obtain by discretely  increasing
  the 4--forms. The answer depends to some degree on the
  compactification radii on the internal space but with modest
  parameters it is not hard to get such a huge number of vacua
  that it is statistically likely to have one in the range
  $\lambda \sim 10^{-120}M_p^4$.

  To see how this works we write the \cc \ as the sum of two terms,
 is the \cc \ for vanishing 4--form, and the contribution of the
 4--forms,
 \be
\lambda =\lambda_0 + \sum_{i=1}^N c_iN_i^2.
 \ee
With a hundred terms and modestly small values for the $c_n$ it is
highly likely to find a value of $\lambda$ in the observed range.
Note that no fine tuning is required, only a very large number of
ways to make the vacuum energy.

The problem with \cite{joe} was clearly recognized by the authors;
The starting point is so far from the \sms \ that none
of the usual tools of approximate supersymmetry are available to
control the approximation. The example was intended only as a model of
what might happen because of the large number of possibilities.

More recently Kachru, Kallosh, Linde and Trivedi \cite{andrei} have improved the
situation by finding an example which is more under control. These
authors subtly use the various ingredients of string theory
including fluxes, branes, anti--branes and instantons to construct
a rather tractable example
 with a small positive \cc .

In addition to arguing that string theory does have many vacua
with positive \cc \ the argument in \cite{andrei} tends to dispel
the idea that vacua, not on the \sms , must have vanishing \cc .
In other words there is no evidence in string theory  that a hoped for but
unknown mechanism
 will automatically force the \cc \ to zero.
It
seems very likely that all of the non--supersymmetric vacua have
finite $\lambda$.

The vacua in \cite{andrei} are not at all simple. They are
jury-rigged, Rube Goldberg contraptions that could hardly have
fundamental significance. But in an anthropic theory simplicity
and elegance are not considerations. The only criteria for
choosing a vacuum is utility, i.e. does it have the necessary
elements such as galaxy formation and complex chemistry that are
needed for life. That together with a cosmology that guarantees a
high probability that
 at least one large patch of space will form
with that vacuum structure is all we need.

\section{The Trouble with de Sitter Space }

The classical  vacuum solution of Einstein's equations with a
positive \cc \ is de Sitter space. It is doubtful that it has a
precise meaning in a quantum theory such as string theory
\cite{naureen, james, lisa}. I want to review some of the reasons
for thinking that \des \ is at best a metastable state.

It is important to recognize that there are two
very different ways to think about \des .  The first is to take a
global view of the spacetime. The global geometry is described by
the metric
\be
ds^2 =R^2 \left\{
dt^2 -(\cosh{t})^2 d^2\Omega_3
\right\}
\ee where
$d^2\Omega_d$ is the metric for a unit d--sphere and $R$ is related to the \cc \ by
\be
R=(\lambda G)^{-\12}.
\ee

Viewing \des \ globally would make sense  if it were a system that could
be studied from the outside by a ``meta--observer". Naively, the
meta--observer would make use of a (time dependent) Hamiltonian to
evolve the system from one time to another. An alternate
description would use a Wheeler de Witt formalism to define a wave
function of the universe on global space--like slices.

The other way of describing the space is the {\it{causal patch}}
or ``Hot Tin Can"
description. The relevant metric is
\be
ds^2 =R^2 \left\{
(1-r^2)dt^2- {1\over (1-r^2)}dr^2 -r^2 d^2\Omega_2.
\right\}
\ee
In this form the metric is static and has a form similar to that
of a black hole. In fact  the geometry has a horizon at $r=1$. The
static patch does not cover the entire global \des \ but is
analogous to the region outside a black hole horizon. It is the
region  which can receive signals from, and send signals to, an
observer located at $r=0$. To such an observer \des \ appears to
be a spherical cavity bounded by a horizon a finite distance away.

Experience with black holes has taught us to be very wary of
global descriptions when horizons are involved. In a black hole
geometry there is no global conventional quantum description of
both sides of the horizon. This suggests that a conventional quantum
description of \des \
  only makes sense within a given observer's
causal patch. The descriptions in different causal patches are
complementary \cite{stretch, whiting} but can not be put together
into a global  description without somehow modifying the rules of quantum
mechanics.

As in the black hole case, a horizon implies a
thermal behavior with a temperature and an entropy. These are
given by
\bea
T\eq {1\over 2\pi R} \cr
S \eq {\pi R^2 \over G }.
\eea
For the rest of this section I will be assuming the causal patch
description of some particular observer.

If the observed ``dark energy" in the universe really is a small positive \cc \  the ultimate future of our universe will be
eternal \des . This would mean not that the future is totally
empty space but that the world will have all the features of an
isolated
finite thermal cavity with finite temperature and entropy. Thermal
equilibrium for such a system is not completely featureless. On
short time scales not much can be expected to happen but on very
long time scales everything happens. A famous example involves a
gas of molecules in a sealed room. Imagine that we start all the
molecules in one corner of the room. In a relatively short time
the gas will spread out to fill the room and come to thermal
equilibrium. During the approach to equilibrium interesting
dissipative structures such as droplets, eddies and vortices form
and then dissipate. The usual assumption is that nothing happens
after that. The entropy has reached its maximum value and the second law
forbids any further interesting history. But on a sufficiently long time scale, large
fluctuations will occur. In fact the phase point will return over
and over to the neighborhood of any point in phase space including
the original starting point. These {\it{Poincare recurrences }}
generally occur on a time scale exponentially large in the thermal
entropy of the system. Thus we define the Poincare recurrence time

\be
T_r=\exp{S}.
\ee
On such long time scale the second law of thermodynamics will
repeatedly be violated by large scale fluctuations.

Thus even a  pure \des \ would have an interesting cosmology of
sorts. The causal  patch of any observer would undergo Poincare recurrences
in which it would endlessly fluctuate back to a state similar to its
starting point, but each time slightly different.

The trouble with such a cosmology is that it relies on very rare
``miracles" to start it off each time. But there are other
miracles which could occur and lead to anthropically acceptable
worlds  with a vastly larger probability than our world.
Roughly speaking the relative  probability of a fluctuation leading to a
given configuration is proportional to the exponential of its
entropy. An example of a configuration far more likely than our
own  would be a world in which everything would be just like our
universe except the temperature of the cosmic microwave background
was ten degrees instead of three. When I say everything is the
same I am including such details as the abundance of the elements.

Ordinarily such a universe would be ruled out on the grounds that
it would take a huge miracle for the helium and deuterium to
survive the bombardment by the extra photons implied by the higher
temperature. That is correct, a fantastic miracle would be
required, but such miracles would occur far more frequently than
the ultimate miracle of returning to the starting point. This can
be argued just from the fact that a universe at  10 degrees K
 has a good deal more entropy than one at 3 degrees. In
a world based on recurrences it would be overwhelming unlikely
that cosmology could be traced back to something like the
inflationary era without a miraculous reversal of the second law
along the way. Thus we are forced to conclude that the sealed tin
can model of the universe must be incorrect, at least for time
scales as long as the recurrence time.

Another difficulty with an eternal \des \ involves a mathematical
conflict between the symmetry of \des \ and the finiteness of the
entropy \cite{naureen}. Basically the argument is that the
finiteness of the \des \ entropy indicates that the spectrum of
energy is discrete. It is possible to prove that the symmetry
algebra of \des \ can not be realized in a way which is consistent
with the discreteness of this spectrum. In fact this problem is
not independent of the issues of recurrences. The discreteness of
the spectrum means that there is a typical energy spacing of order
\be
\Delta E \sim \exp{-S}.
\ee
The discreteness of the spectrum can only manifest itself on time
scales of order $(\Delta E)^{-1}$ which is just the recurrence
time. Thus there are problems with realizing the full symmetries
of \des \ for times as long as $T_r$.

Finally another difficulty for eternal \des \ is that it does not
fit at all well with string theory. Generally the only objects in
string theory which are rigorously defined are S--Matrix elements.
Such an S matrix can not exist in a thermal background. Part of
the problem is again the recurrences which undermine the existence
of asymptotic states. Unfortunately there are no known observables
in \des \ which can substitute for S--matrix elements. The
unavoidable implication of the issues I have raised is that
eternal \des \ is an impossibility in a properly defined quantum
theory of gravity.

\setcounter{equation}{0}
\section{de Sitter Space is Unstable}

In \cite{andrei} a particular string theory vacuum with positive $\lambda$ was
studied. One of the many interesting things that the authors found
was that the vacuum is unstable with respect to tunneling to other
vacua. In particular the vacuum can tunnel back to the \sms \ with
vanishing \cc . Using instanton methods the authors
calculated that the lifetime of the vacuum is less than the Poincare
recurrence time. This is no accident. To see why it
always must be so, let's consider the effective potential that the
authors of  \cite{andrei} derived. The only modulus which is  relevant
is the overall size of the compact manifold $\Phi$. The potential is
shown in Figure 1. The de Sitter vacuum occurs at the point
$\Phi=\Phi_0$. However, the absolute minimum of the potential
occurs not at $\Phi_0$ but at $\Phi=\infty$. At this point the
vacuum energy is exactly zero and the vacuum one of the ten
dimensional vacua of the \sms . As was noted long ago by Dine and
Seiberg there are always runaway solutions like this in string
theory. The potential on the \sms \ is zero and so it is always
possible to lower the energy  by tunneling to a point on the \sms.
\begin{figure}[h]
\begin{center}
\leavevmode \epsfbox{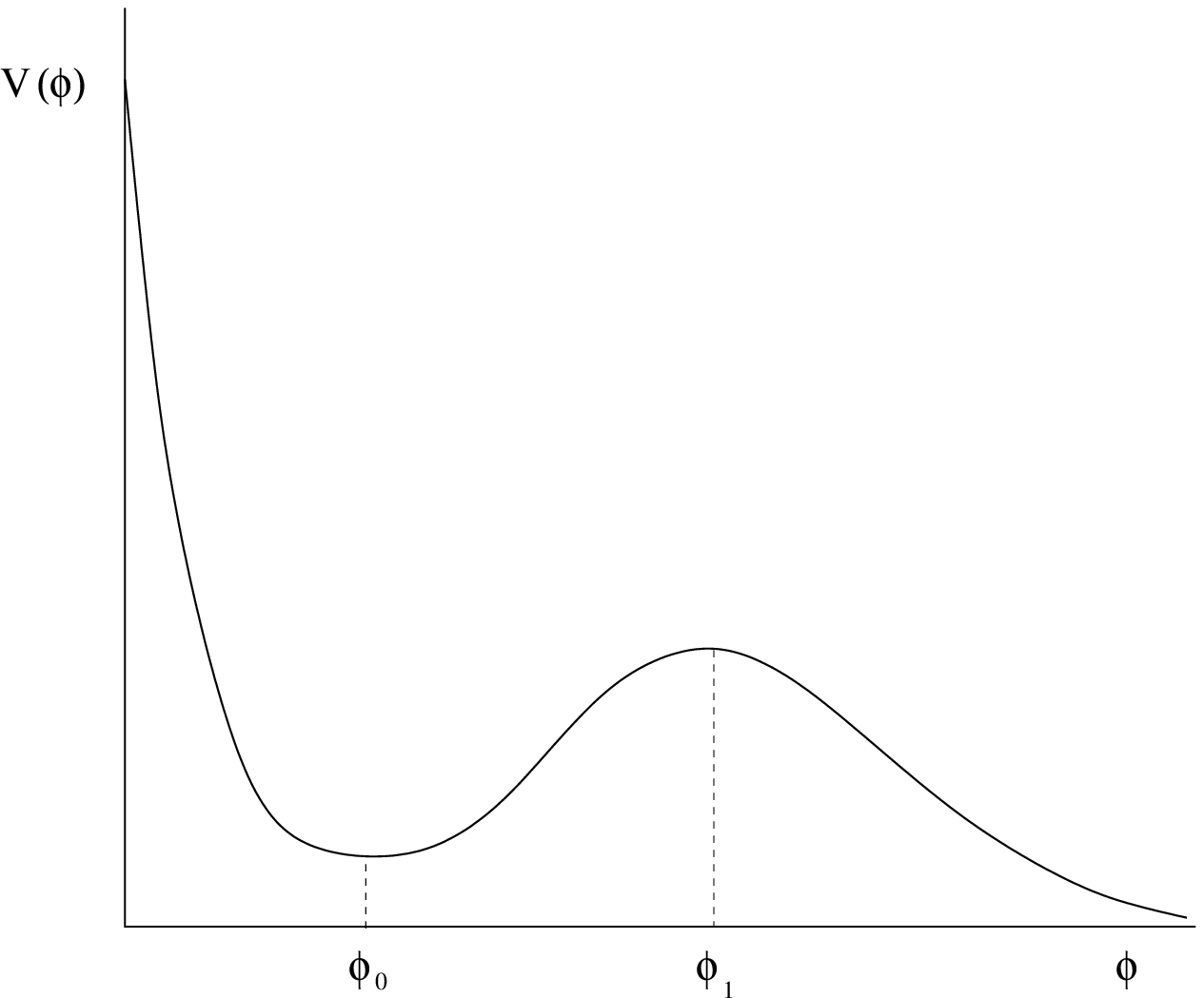} \caption{}
\end{center}
\end{figure}

Suppose we are stuck in the potential well at $\Phi_0$. The vacuum
of the causal patch
 has a finite entropy and fluctuates up and down the
walls of the potential. One might think that fluctuations up the
sides of the potential are Bolzmann suppressed. In a usual thermal
system there are two things that suppress fluctuations. The first
is the Bolzmann suppression by factor
$$
\exp{- \beta E}
$$
and the second is entropy suppression by factor
$$
\exp{S_f -S}
$$
where $S$ is the thermal entropy and $S_f$ is the entropy
characterizing the fluctuation which is generally smaller than
$S$. However in a gravitational theory in which space is bounded
(as in the static patch) the total energy is always zero, at least
classically. Hence the only suppression is entropic. The phase
point wanders around in phase space spending a time in each region
proportional to its phase space volume, i.e. $\exp{-S_f}$.
Furthermore the typical time scale for such a fluctuation to take
place is of order
\be
T_f \sim \exp{S -S_f}.
\ee

Now consider a fluctuation which brings the field $\phi$ to the
top of the local maximum at $\phi=\phi_1$ in the entire causal patch.
The entropy at the top
of the potential is given in terms of the \cc \ at the top. It is
obviously positive and less than the entropy at $\phi_0$. Thus the
time for the field to fluctuate to $\phi_1$ (over the whole causal patch)
is strictly less than the recurrence
time $\exp{S}$. But once the field gets to the top there is no
obstruction to it rolling down the other side to infinity. It follows that a de
Sitter vacuum of string theory is never longer lived than $T_r$
and furthermore we  end up at a supersymmetric point of vanishing \cc .

There are other possibilities. If the cosmological constant is not
very small it may tunnel over the nearest mountain pass to a
neighboring valley of smaller positive \cc. This will also take
place on a time scale which is too short to allow recurrences. By
the same argument it will not stay in the new vacuum indefinitely.
It may find a vacuum with yet smaller \cc \ to tunnel to.
Eventually it will have to make a transition out of the space of
vacua with positive cosmological constants\footnote{Tunneling to
vacua with negative \cc \ may or may not be a possibility. However
such a transition will eventually lead to a crunch--singularity.
Whether the system survives the crunch is not known. It should be
noted that transitions to negative \cc \ are suppressed and can
even be forbidden depending on  magnitudes of the vacuum energies,
and the domain wall tension. I will assume that such transitions
do not occur.}.

\setcounter{equation}{0}
\section{Bubble Cosmology}

To make use of the enormous diversity of environments that string
theory is likely to bring with it, we need a dynamical cosmology
which,
with high probability, will
populate one or more regions of space with an anthropically
favorable vacuum. There is a natural candidate for such a
cosmology that I'll explain from the global perspective.

For simplicity let's temporarily  assume that there are only two
vacua, one with positive \cc \ $\lambda$, and one with vanishing
\cc . Without worrying how it happened we suppose that some region
of the universe has fallen into the minimum with positive \cc .
>From the global perspective it is inflating and new Hubble volumes
are constantly being produced by the expansion. Pick a time--like
observer who looks around and sees a static universe bounded by a
horizon. The observer will eventually observe a transition in
which his entire observable region slides over the mountain pass
and settles to the region of vanishing $\lambda$. The observer
sees the horizon--boundary quickly recede, leaving in its wake an
infinite open Freedman, Robertson, Walker  universe with negative
spatial curvature. The final geometry has light--like and time
like future infinities similar to flat space.
\begin{figure}
\begin{center}
\leavevmode \epsfbox{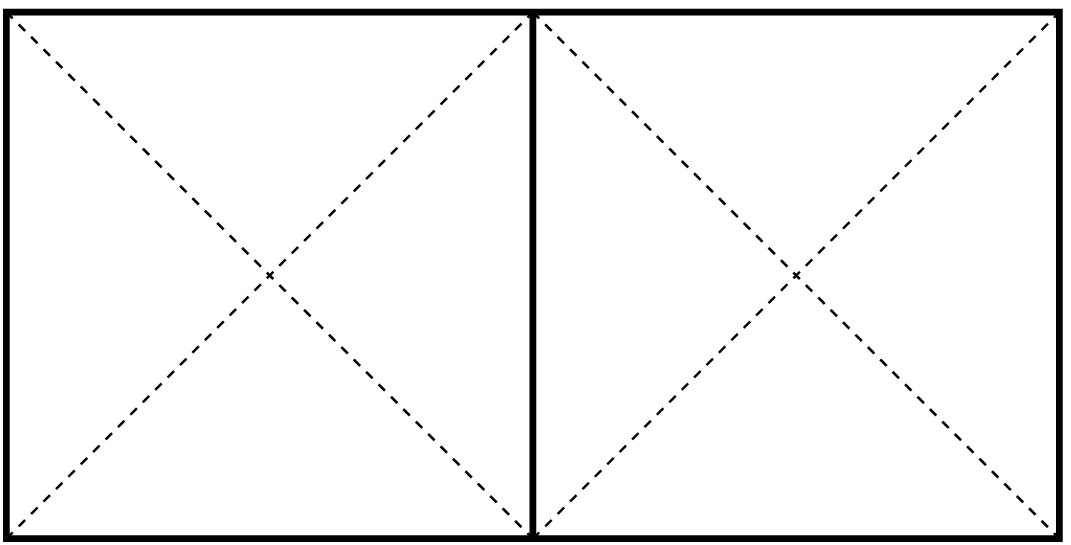} \caption{}
\end{center}
\end{figure}

It is helpful to draw some Penrose diagrams to illustrate the history.
For this purpose we turn to the global point of view.
First draw a diagram representing pure \des . See Figure 2.
The figure also shows two observers whose causal patches
overlap for some period of time.
In Figure 3,  the same geometry is shown except that the
formation of a bubble of $\lambda=0$ vacuum is also depicted. The
bubble is created at point $(a)$ and expands with velocity that
approaches the speed of light. Eventually the growing bubble
intersects the infinite future of the \des \ but the geometry
inside the bubble continues and forms a future infinite null.
Notice that even though the observer's final universe is
infinite his past light cone does not include the whole global
space--time. In fact his causal diamond is not much bigger than it
would have been if the space never decayed.
\begin{figure}
\begin{center}
\leavevmode \epsfbox{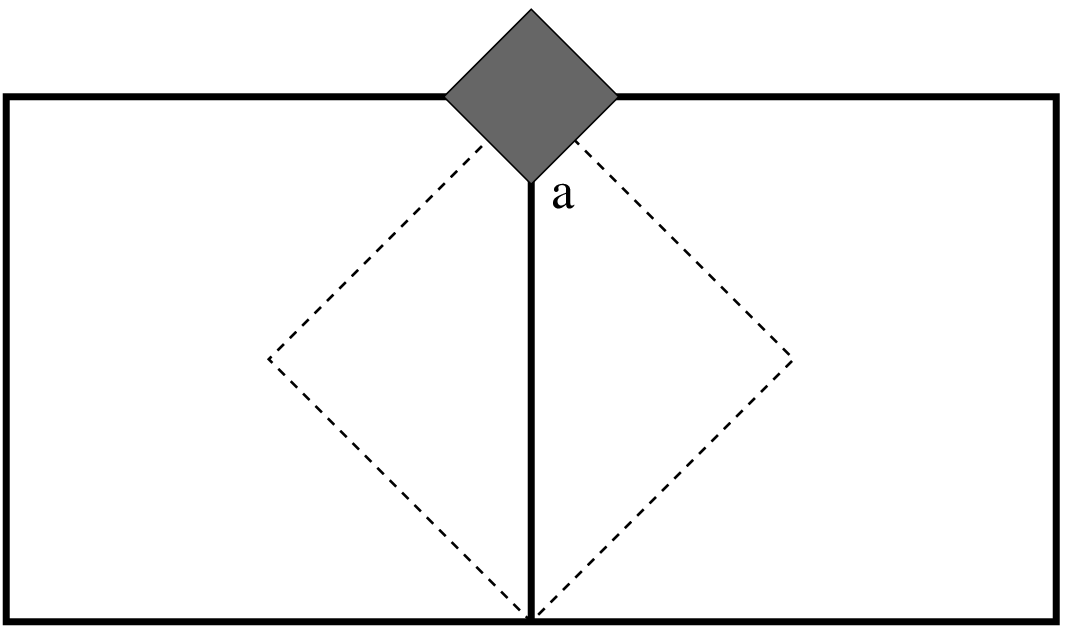} \caption{}
\end{center}
\end{figure}

The region outside the bubble is still inflating and disappearing
out of causal contact with the observer. From the causal patch
viewpoint the entire world has been swallowed by the bubble.

Now let us take the more global view. The bubble does not swallow
the entire global space but leaves part of the space still
inflating.
Inevitably bubbles will
form in this region. In fact if we follow the world line of any
observer, it will eventually be swallowed by a bubble of $\lambda=0$
vacuum. The line representing the romote future in Figure 3 is
replaced by a jagged fractal as in Figure 4. Any observer
eventually ends up at the top of the diagram in one of
infinitely many time--like infinities. This process
leading to infinitely many disconnected  bubble universes
is essentially  similar to the process of eternal inflation
envisioned by Linde.
\begin{figure}
\begin{center}
\leavevmode \epsfbox{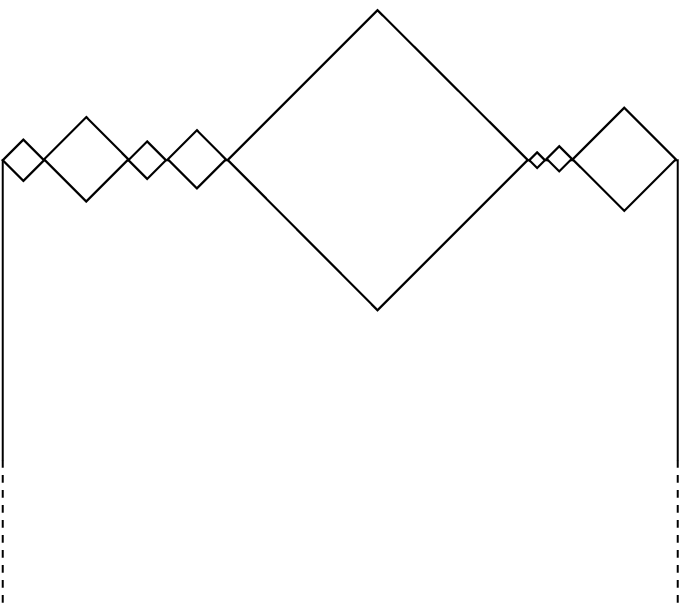} \caption{}
\end{center}
\end{figure}

The real landscape is not comprised of only two vacua. If an observer
starts with a large value of the \cc \ there
will be many ways for the causal patch to desend to the \sms. From
the global viewpoint a bubbles will form in  neighboring valleys
with  somewhat smaller \cc . Since each bubble has a positive \cc \ it will be
inflating but the space between bubbles is
inflating faster so the bubbles go out of causal contact with one
another. Each bubble evolves in isolation from all the others.
Furthermore, in a time too short for recurrences, bubbles will
nucleate within the bubbles. Following a single observer within
his own causal patch, the cosmological constant decreases in a
series of events until the causal patch finds itself in
the \sms . Each observer will see a series of vacuums decending
down to the \sms \ and the chances that he passes through an
anthropically acceptable vacuum is most likely very small. But on
the other hand the global space contains an infinite number of
such histories and some of them will be acceptable.

The only problem with the cosmology that I just outlined is that
it is formulated in global coordinates. From the viewpoint of any
causal patch, all but one of the bubbles is outside the horizon.
As I've emphasized, the application of the ordinary rules of
quantum mechanics only makes sense within the horizon of an
observer. We don't know the rules for putting together the various
patches into one comprehensive
global description and until we do there can not be any firm basis
for the kind of anthropic cosmology I described. Nevertheless the
picture is tempting.

\setcounter{equation}{0}
\section{Cosmology as a Resonance}

The idea of scalar fields and potentials is approximate once we
leave the \sms . So is the notion of a stable de Sitter vacuum.
The problem is familiar. How do we make precise sense of an
unstable state in quantum mechanics. In ordinary quantum mechanics
the clearest situation is when we can think of the unstable state
as a resonance in a set of scattering amplitudes. The parameters
of a resonance, i.e. its width and mass are well defined and don't
depend on the exact way the resonance was formed. Thus even black
holes have precise meaning as resonant poles in the S--matrix.
Normally we can not compute the scattering amplitudes that
describe the formation and evaporation of a black hole but it is
comforting that an exact criterion exists.

 In the case of a black
hole the density of levels is enormous being proportional to the
exponential of the entropy. The spacing between levels is
therefore exponentially small.  On the other hand the width of
each level is not very small. The lifetime of a state is the time
that it takes to emit a single quantum of radiation and this is
proportional to the Schwarzschild radius. Therefore the levels are
broadened by much more than their spacing. The usual resonance
formulas are not applicable but the precise definition of the
unstable state as a pole in the scattering amplitude is. I think
the same things can be said about the unstable de Sitter vacua but
it can only be understood by returning to the causal patch way of
thinking. Therefore let's focus on the causal patch of one
observer. We have discussed the observer's future history and
found that it always ends in an infinite expanding supersymmetric
open Freedman universe. Such a universe has the usual kind of
asymptotic future consisting of time--like and light--like
infinity. There is no temperature in the remote future and the
geometry permits particles to separate and propagate as free
particles just as in flat space--time.

Now let's consider the observer's past history. The same argument which
says that the observer will eventually make a transition to
$\lambda=0$ in the far future can be run backward. The observer could only
have gotten to the de Sitter vacuum by the time--reversed history
so he must have originated from a collapsing open universe. The
entire history is shown in Figure 5. The history may seem
paradoxical since it requires  the second law of thermodynamics to
be violated in the past. A similar paradox arises in a more
familiar setting. Let me return to the sealed room filled with gas
molecules except that now one of the walls has a small hole that
lets the gas escape to unbounded space. Suppose we find the gas
filling the room in thermal equilibrium at some time. If we run
the system forward we will eventually find all the molecules have
escaped and are on their way out, never to return. But it is also
true that if we run the equations of motion backwards we will
eventually find all the molecules outside the room moving away.
Thus the only way the starting configuration could have occurred
is if the original molecules were converging from infinity toward
the small hole in the wall.
\begin{figure}
\begin{center}
\leavevmode \epsfbox{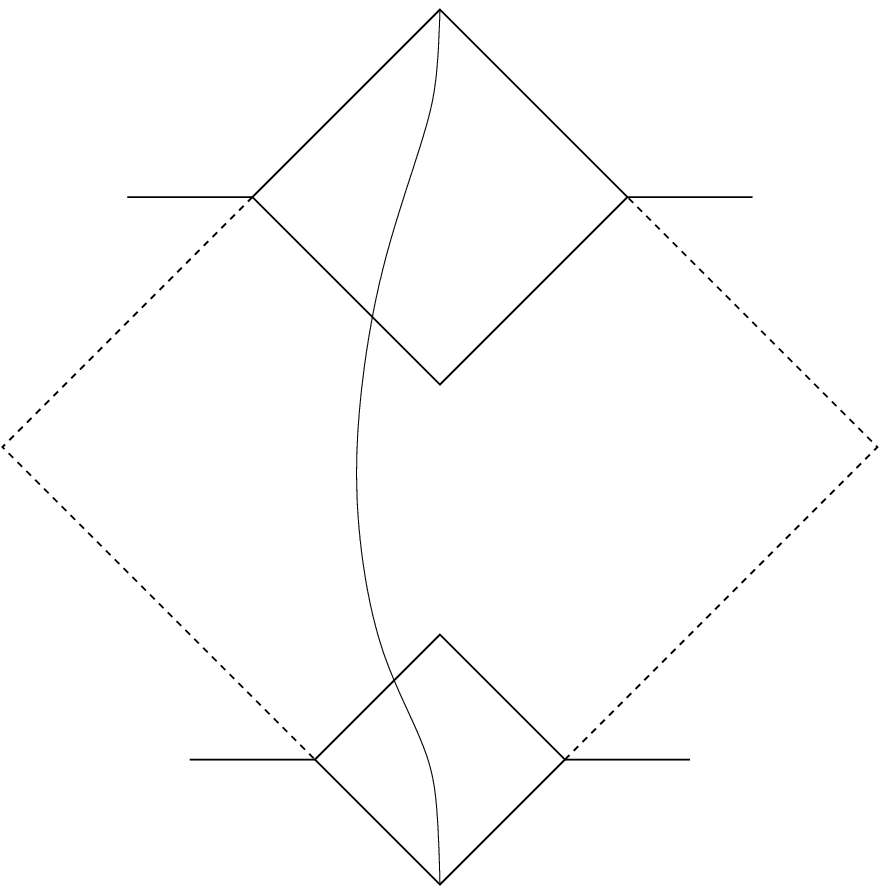} \caption{}
\end{center}
\end{figure}

If we are studying the system quantum mechanically, the metastable
 configuration with all the molecules in the room would be an
 unstable $resonance$ in a scattering matrix describing the many
 body scattering of a system of molecules with the walls of the
 room. Indeed the energy levels describing the molecules trapped
 inside the room are complex due to the finite lifetime of the
 configuration.

 This suggests a view of the  intermediate \des \ in Figure 5 as an
 unstable resonance in the scattering matrix connecting states in
 the asymptotic $\lambda =0$ vacua. In fact we can estimate the
 width of the states. Since the lifetime of the \des \ is always
longer than the recurrence time, generally by a huge factor, the
width $\gamma$ satisfies
$$
\gamma >> \exp{-S}.
$$
On the other hand the spacing between levels, $\Delta E$, is of
order $\exp{-S}$. Therefore
\be
\gamma>> \Delta E
\ee
so that the levels are very broad and overlapping as for the black hole.

No perfectly precise definition exists in string theory for the
moduli fields or their potential when we go away from the \sms .
The only precise definition of the de Sitter vacua seems to be as
complex poles in some new sector of the scattering matrix between
states on the \sms .

Knowing that a black hole is a resonance in a scattering amplitude
does not tell us much about the way real black holes form. Most of
the possibilities for black hole formation are just the time
reverse of the ways that it evaporate. In other words the
overwhelming number of initial states that can lead to a black
hole consist of thermal radiation.  Real black holes
in our universe form from stellar collapse which is just one
channel in a huge collection of S--matrix ``in states". In the
same way the fact that cosmological states may be thought of in a
scattering framework is in itself does not shed much light on the
original creation process.

\setcounter{equation}{0}
\section{Conclusion}

Vacua  come in two varieties, supersymmetric and otherwise. Most
likely the non--supersymmetric vacua do not have vanishing \cc \
but it is plausible that there are so many of them that they
practically form a continuum. Some tiny fraction have \cc \ in the
observed range. With nothing preferring one vacuum over another,
the anthropic principle comes to the fore whether or not we like
the idea. String theory provides a framework in which this can be
studied in a rigorous way. Progress can certainly be made in
exploring the landscape. The project is in its infancy but in time
we should know just how rich it is. We can argue the philosophical
merits of the anthropic principle but we can't argue with
quantitative information about the number of vacua with each
particular property such as the \cc , Higgs mass or fine structure
constant. That information is there for us to extract.

Counting the vacua is important but not sufficient. More
understanding of cosmological evolution is essential to
determining if the large number of possibilities are realized as
actualities. The vacua in string theory  with $\lambda
>0$ are not stable and decay on a time scale smaller than the
recurrence time. This is very general and also very fortunate
since there are serious problems with stable de Sitter space.

 The
instability also allows the universe to sample all or a large part
of the landscape by means of bubble formation.  In such a world
the probability that some region of space has suitable conditions
for life to exist can be large.

 The bubble universe based on Linde's eternal
inflation seems promising but it is unclear how to think about it
with precision. There are real conceptual problems having to do
with the global view of spacetime. The main problem is to
reconcile two pictures; the causal patch picture and the global
picture. String theory has provided a testing ground for some
important relevant  ideas such as black hole complementarity
\cite{stretch, whiting} and the Holographic principle
 \cite{hooft,holo}. Complementarity requires the observer's side of the
horizon to have a self contained conventional quantum description.
It also prohibits a conventional quantum description that covers
the interior and exterior simultaneously. Any attempt to describe
both sides as a single quantum system will  come into conflict
with one of three sacred principles \cite{birthday}. The first is
the equivalence principle which says that a freely falling
observer passes  the horizon without incident. The second says
that experiments performed outside a black hole should be
consistent with the rules of quantum mechanics as set down by
Dirac in his textbook. No loss of quantum information should take
place and the time evolution should be unitary. Finally the rules
of quantum mechanics forbid information duplication. This means
that we can not resolve the so called information paradox by
creating two copies (quantum Xeroxing) of every bit as it falls
through the horizon; at least not within the formalism of
conventional quantum mechanics. The complementarity and
holographic principles have been convincingly confirmed by the
modern methods of string theory \cite{juan}. The inevitable
conclusion is that a global description of geometries with
horizons, if it exists at all, will not be based on the standard
quantum rules.

Why is this important for cosmology? The point is that the eternal
inflationary production of an infinity of bubbles takes place
behind the horizon of any given observer. It is not something that
has a description within one causal patch. If it makes sense, a
global description is needed  but if cosmic event horizons are at
all like black hole horizons then any global description will
involve wholly new elements. If I were to make a wild guess about
which rule of quantum mechanics has to be given up in a global
description of either black holes or cosmology I would guess it is
the Quantum Xerox Principle \cite{birthday}. I would look for a
theory which formally allowed quantum duplication but cleverly
prevents any observer from witnessing it. Perhaps then the
replication of bubbles can be sensibly described.

Progress may also be possible in sharpening the exact mathematical
meaning of the de Sitter vacua. Away from the \sms , the concept
of a local field  and the  effective potential is at best
approximate in string theory. The fact that the vacua are false
metastable states makes it even more problematic to be precise. In
ordinary quantum mechanics the best mathematical definition of an
unstable state is as a resonance is   amplitudes for scattering
between very precisely defined asymptotic states. Each metastable
state corresponds to pole whose real and imaginary parts define
the energy and inverse lifetime of the state.

I have argued that each causal patch begins and ends with an
asymptotic ``roll" toward the \sms . The final state have the
boundary conditions of an FRW open universe and the initial states
are time reversals of these. This means we may be able to define
some kind S--matrix connecting initial and final asymptotic
states. The various intermediate metastable de Sitter phases would
be exactly defined as resonant resonances in this amplitude.

At first this proposal sounds foolish. In general relativity
initial and final states are very different. Black holes make
sense. White holes do not.  Ordinary things fall into black holes
and thermal radiation comes out. The opposite never happens. But
this is deceiving. Our experience with string theory has made it
clear that the fundamental micro--physical input is completely
reversible and that black holes are most rigorously defined in
terms of resonances in scattering amplitudes \footnote{The one
exception is black holes in anti--de Sitter space which are
stable. }. Of course knowing that a black hole is an intermediate
state in a tremendously complicated scattering amplitude does not
really tell us much about how real black holes form. For that we
need to know about stellar collapse and the like. But it does
provide an exact mathematical definition of the states that
comprise the black hole ensemble.

To further illustrate the point let me tell a story:

 Two future
astronauts in the deep empty reaches of  outer space   discover a
sealed capsule. On further inspection they find a tiny pin--hole
in the capsule and air is slowly leaking out. One says to the
other, ``Aha, we have discovered an eternal air tank. It must have
been here forever." The other says ``No, you fool. if it were here
forever the air would have leaked out long ago (infinitely long
ago)." So the first one thinks and says, ``Yes, you are right.
Let's think. If we wait long enough, all the air will be streaming
outward in an asymptotic final state. That is clear. But,  because
of micro--reversibility,  it is equally clear that if we go far
into the past that all the air must have been doing the reverse.
In fact the  quantum states with air in the capsule are just
intermediate resonances in the scattering of a collection of air
molecules with the empty capsule." The second astronaut looks at
him as if he were nuts. ``Don't be a dope. That's just too
unlikely. I guess someone else was here not so long ago and filled
it up."

Both of them can be right. The quantum states of air in a tank are
mathematical resonances in a scattering matrix. And it may also be
true that the laws of an isolated system of gas and tank may have
been temporarily interfered with by another presence. Or we might
say that the scattering states need to include not only air and
tank but also cosmonauts and their apparatus.

It's in this sense that I propose that \des \ can be
mathematically defined in terms of singularities in some kind of
generalized S--matrix. But in so doing, I am not really telling
you much about how it all started.

>From the causal patch viewpoint the evolutionary endpoints seems
to be an approach to some point on the \sms . After the last
tunneling the universe enters an final open FRW expansion toward
some flat supersymmetric solution. This is not to be thought of as
a unique quantum state but as a large set of states with similar
evolution. Running the argument backward (assuming microscopic
reversibility) we expect the initial state to be the time reversal
of one of the many future endpoints. We might even hope for a
scattering matrix connecting initial and final states. de Sitter
minima would
 be an enormously large density of complex poles in the amplitude.

 One last point: The final and initial states do not have to be
 four dimensional. In fact in the example given in \cite{andrei},
 the modulus describing the overall size of the compact space
 roles to infinity, thus creating a ten or possibly and eleven
 dimensional universe.

 \setcounter{equation}{0}
\section{Acknowledgements}

I am indebted  to
 Simeon Hellerman, Ben Freivogel, Renata Kallosh, Shamit
Kachru, Eva Silverstein and Steve Shenker. Special thanks goes to
Matt Kleban and Andrei Linde for explaining many things to me.

\end{document}